\documentclass{article}

\usepackage{arxiv}

\usepackage[utf8]{inputenc} 
\usepackage[T1]{fontenc}    
\usepackage{hyperref}       
\usepackage{url}            
\usepackage{booktabs}       
\usepackage{amsfonts}       
\usepackage{nicefrac}       
\usepackage{microtype}      
\usepackage{lipsum}		
\usepackage{graphicx}
\usepackage{natbib}
\usepackage{doi}
\usepackage{amsmath} 
\usepackage{tabularx}

\title{Agent-based macroeconomics for the UK’s Seventh Carbon Budget}

\author{{\hspace{1mm}Tom Youngman}\\
	Institute for New Economic Thinking\\
	\texttt{thomas.youngman@inet.ox.ac.uk} \\
	\And
	{\hspace{1mm}Tim Lennox} \\
	Department for Energy Security and Net Zero\\
	\And
    {\hspace{1mm}M Lopes Alves} \\
	Institute for New Economic Thinking\\
	\And
	{\hspace{1mm}Pirta Palola} \\
	Institute for New Economic Thinking\\
	\And
	{\hspace{1mm}Brendon Tankwa} \\
	Institute for New Economic Thinking\\
	\And
	{\hspace{1mm}Emma Bailey} \\
	Institute for New Economic Thinking\\
	\And
	{\hspace{1mm}Emilien Ravigné} \\
	Institute for New Economic Thinking\\
	\And
	{\hspace{1mm}Thijs Ter Horst} \\
	Institute for New Economic Thinking\\
	\And
	{\hspace{1mm}Benjamin Wagenvoort} \\
	Institute for New Economic Thinking\\
    \And
	{\hspace{1mm}Harry Lightfoot Brown} \\
	Department for Energy Security and Net Zero\\
    \And
	{\hspace{1mm}José Moran} \\
	Macrocosm\\
    \And
	{\hspace{1mm}Doyne Farmer} \\
	Institute for New Economic Thinking\\
}

\date{}

\renewcommand{\undertitle}{To the presented at the Workshop of Macroeconomic Implications of Climate-Related Risks: Challenges and Opportunities for the Low-Carbon Transition}

\hypersetup{
pdftitle={Agent-based macroeconomics for the UK’s Seventh Carbon Budget},
pdfsubject={cs.OH},
}

\begin{document}
\maketitle

\begin{abstract}
In June 2026, the UK government will set its carbon budget for the period 2038 to 2042, the seventh such carbon budget (CB7) since the Climate Change Act became law in 2008. For the first time, this carbon budget will be accompanied by a macroeconomic assessment of its impact on growth, employment, inflation and inequality. Researchers from the Institute of New Economic Thinking (INET) Oxford are working in partnership with the Department for Energy Security and Net Zero to deliver this assessment using our data-driven macroeconomic agent-based model (ABM).\\
This extended abstract presents the work in progress towards this pioneering policymaking using our data-driven macroeconomic ABM. We are conducting our work in three work packages. By the time of the workshop, we hope to be able to present preliminary findings from the first two work packages.\\
In WP1, we adapt an existing macro-ABM prototype and build a UK macroeconomic baseline. The main task for this is initialising the model with suitable UK household microdata. We present the options considered and the approach settled upon. In WP2, we conduct preliminary modelling that represents UK decarbonisation as an external shock to financial flows and technical coefficients. In order to present results in time to influence the June 2026 policy decision, this second work package exogenously forces the ABM to follow the CB7 green investment and associated technological change projections provided by the Climate Change Committee. Finally, we will implement more sophisticated social and technological learning packages in WP3, building our own projections of likely decarbonisation pathways that may diverge from UK government plans. For the workshop, we will present the progress of WP1 and WP2.
\end{abstract}


\section*{Policy background}

In June 2026, the UK government will set its seventh carbon budget (CB7) for the period 2038 to 2042. The UK carbon budgets are five-year caps on territorial emissions levels set under the Climate Change Act 2008. The first six UK carbon budgets were set without quantitative macroeconomic or inequality analysis. As easier mitigation options get exhausted, climate policies on the agenda require more widespread social and technological change. The UK's Climate Change Committee (CCC) foresee the CB7 requiring additional investment in the region of a percentage point of GDP, and major changes in practices and technologies within many industries \citep{ccc2025seventh}. To successfully navigate such major changes through potential social and political opposition, macroeconomic and distributional analysis can no longer be overlooked. The Department for Energy Security and Net Zero (DESNZ) agree, and have identified macroeconomic and distributional analysis as evidence gaps that they must address to set the CB7. We are working in partnership with DESNZ to fill that gap with a state-of-the-art, data-driven agent-based macroeconomic model. For the first time, this seventh carbon budget will be accompanied by a macroeconomic assessment of its impact on growth, employment, inflation and inequality. This paper presents our work in progress towards this pioneering policy-making use of our macroeconomic ABM.

\section*{INET Oxford's macroeconomic model}

Data-driven macroeconomic ABMs are emerging as a new economic modelling approach and have already demonstrated the potential to rival the forecasting performance of DSGE models \citep{poledna2023forecasting}. ABMs are a class of computational models that simulate a system of interacting agents, making it possible to study how macrolevel phenomena arise from microlevel interactions. Macroeconomic ABMs simulate macroeconomic aggregates from interactions between individuals, households and firms. Data-driven macroeconomic ABMs use real survey microdata to define their synthetic populations of households and firms \citep{poledna2023forecasting,wiese2024forecasting}.

The data-driven ABM developed at INET Oxford is a prototype model that covers all OECD countries \citep{wiese2024forecasting}. Our ABM has the necessary characteristics for modelling the carbon budgets' macroeconomic and inequality impacts. The ABM framework allows easier modelling of a wider range of socioeconomic policy interventions, in contrast to most macroeconomic climate modelling that flattens all policies into a carbon tax equivalent. As the ABM outputs individual income and wealth balance sheets for each of the thousands or millions of agents represented in each time period, it has great potential to conduct dynamic analyses of the distributional impacts of policy over time. The ABM also has the flexibility to represent constraints that shape real-world sustainability transitions, such as delays between the purchase and deployment of capital goods.

A complete model specification is provided by \citet{wiese2024forecasting} and is currently in the peer review process.

\section*{1. UK baseline}

There are also precedents of linking the Wealth and Assets Survey with the Living Costs and Food Survey to generate household-level greenhouse gas emissions estimates \citep{buchs2024emission}. The PolicyEngine microsimulation model addresses undersampling of rich households by using the Wealth and Assets Survey to impute additional wealth households into the Family Resources Survey \citep{policyengine2025data}.

Once we initialise the ABM with UK household microdata, we will rerun our simulation-based inference (SBI) process to calibrate the small subset of model parameters that are difficult to infer from real world data. This process is detailed in the working paper describing the model \citep{wiese2024forecasting}.

\section*{2. External shock}

This provisional analysis assumes that Climate Change Committee (CCC) assessments of green investment needs and the potential for technological progress are correct \citep{ccc2025seventh}. We model the CB7 policy implications against a 'no further decarbonisation action' baseline as defined by the CCC.

Like \citet{poledna2023forecasting}, the lack of a bond market is for now a weakness in our model, so we need to be careful to ensure government debt stays within realistic ranges, or add provisions to increase the cost of government borrowing as debt increases.

\begin{table}[t]
\caption{How policy levers combine into scenarios}
\centering
\small
\begin{tabularx}{\linewidth}{lXXXX}
\toprule
 & Households & Firms & Government & Central Bank \\
\midrule
Financial flow 
& B: Forced to spend on green capital and goods 
& A: Forced to spend on green capital and goods 
& C: Subsidies to firms or households, financed via fiscal expansion or reduced spending 
& Implicitly sets government borrowing constraint \\

Technology change 
& E: Consumption weights change 
& D: Intermediate consumption changes 
& -- 
& -- \\
\bottomrule
\end{tabularx}
\label{tab:table}
\end{table}

\section*{Future Work}
\section*{3. Learning}

In this final work package, we will incorporate technological learning into the ABM firm environment. We will not have results available for this work package by the time of the workshop in February, but we hope sharing our plans for the spring and summer gives rise to helpful suggestions and feedback.

INET Oxford researchers have developed techniques that use historical data to forecast deployment and costs of emerging technologies \cite{baumgartner2025renewable, way2022technology}. These methods rely primarily on S-curve adoption dynamics and Wright's Law cost learning. We embed this S-curve modelling directly into the macroeconomic framework, allowing technology costs to fall endogenously as deployment expands. For renewable power, we will draw the feasible ranges of technological improvements from INET's work on wind and solar, which suggests that global wind costs will steadily approach a floor of around 35USD/MWh, reaching 43USD/MWh in 2050, whereas global solar costs will continue to decline exponentially, falling to $3\text{--}15$ USD/MWh in 2050 and potentially even lower later in the century \cite{baumgartner2025renewable}. We will also draw on INET research on how technology deployment patterns vary based on national contexts, with global cost trends exerting strong influence on national solar costs, but considerably less so for wind \cite{tankwa2025technological, tankwa2025renewable}.

The structure of the ABM allows for incorporating social learning and findings from behavioural economics. For example, there is evidence that take-up of heat pumps and solar panels follows social networks, with deployment accelerating around local hotspots \cite{bartonhenry2021decay}. In future work, we plan to implement a specific decision process for the purchase of household consumer durables like heat pumps and electric vehicles. These purchases will change consumers' consumption weights for other goods like electricity and natural gas. Household nodes will be linked in a network according to geographical and social criteria, along which we will allow influence to spread in the process of making decisions about decarbonising consumer durables.

\section*{Conclusion}

This extended abstract outlines the ongoing work toward a macroeconomic assessment of the UK's CB7 using a data-driven macroeconomic ABM. The project is oriented toward policy-making and aims to influence the UK government's CB7 decision, scheduled for June 2026.

The first milestone is to deliver an initial analysis of external shocks on financial flows and technical coefficients in the UK economy that correspond to CB7 by mid-February. The model development will continue through June, allowing for the exploration of dynamics and improvements to the codebase's modularity.

Beyond influencing the CB7 guidelines, our project seeks to contribute more broadly to the case for ABMs as practical tools for better policy analysis. Our goal is to show how ABMs can complement existing modelling tools, particularly in contexts involving structural change such as the transition to net zero.

\bibliographystyle{unsrtnat}
\bibliography{references}

\end{document}